\documentclass[seceq]{ptptex}
\usepackage{wrapft}

\notypesetlogo  

\title{Adopting the Uncertainty Principle for the Entropy Estimation of \\
\vspace{0.5cm}

Black Holes, de Sitter Space and Rindler Space

\vspace{0.5cm}

}

\author{
Tetsuya \textsc{Hara}\footnote{E-mail: hara@cc.kyoto-su.ac.jp}, 
 Keita \textsc{Sakai},
and Daigo \textsc{Kajiura}\footnote{E-mail: i253037@cc.kyoto-su.ac.jp}
}

\inst{Department of Physics, Kyoto Sangyo University, Kyoto 603-8555, Japan
}

\abst{
By a simple physical consideration and uncertain principle, we derive that temperature is proportional 
to the surface gravity and 
entropy is proportional to the surface area of the black hole. We apply the same consideration to 
de Sitter space and estimate the temperature and entropy of the space, 
then we deduce that the entropy is proportional to the boundary surface area. By the same consideration, 
we estimate the temperature and entropy in the uniformly accelerated system (Rindler coordinate).
The cases in higher dimensions are considered. 
}

\begin{document}
\maketitle

\section{Introduction}

Although it has passed almost 30 years since the discussion of the thermodynamics of black hole began, 
there are still many problems about the fundamental concepts.  \cite{rf:BEK,rf:JAC,rf:HAW1} 
But several striking concepts are accepted. One of them is that the entropy of the black hole is proportional to 
the surface which characterizes the thermodynamics of black hole. \cite{rf:BEK,rf:BEK2} Such statements have 
been discussed under the quantum field theory \cite{rf:BIR, rf:CARL} and  superstring theory. \cite{rf:HOO, rf:MAR, rf:POL}

Some results could be understood if heuristic assumptions are adopted. One of the key 
concept is the uncertainty principle. \cite{rf: SCA} As the black hole has the gravitational radius, there is the corresponding 
momentum. If the black hole is formed by such photons or particles corresponding to this momentum, 
it could be speculated that the temperature of the black hole is proportional to the surface gravity and 
it could be explained that the entropy of the black hole is proportional to the surface of the black hole.

Recent Type Ia supernovae and WMAP observations confirm the existence of cosmological constant. \cite{rf:PER, rf:RIE, rf:SPE} In the space
 with the cosmological constant (de Sitter space), each observer is surrounded by the 
event horizon of the cosmological scale 
and it is discussed that the entropy of the universe is proportional to the horizon area. \cite{rf:PAD1} 
In this case there is the characteristic length derived from the cosmological constant, 
so it could be speculated the temperature of the horizon from the uncertainty principle. 
If we assume that the universe is composed of particles with such de Brogile length,
it could be derived that the entropy of the universe is related to the surface of the horizon.

The similar consideration is applied to the uniformly accelerated coordinate (Rindler coordinate), where it has 
the characteristic length due to its acceleration. If we assume that the acceleration is due to the gravity of 
the gravitational mass of particles,  
it is derived that the entropy per surface is constant. 

The detailed derivations about the black hole, de Sitter space and Rindler space are given in $\S$ 2, 3 and 4, 
respectively. The cases in higher dimensions are considered in $\S$ 5.  The conclusions and problems are discussed in $\S$ 6.
 In the following, it is an order of magnitude argument and units with $c=\hbar=G=k_B=1$ are used, 
however in many cases the physical constants are expressed.

\section{Entropy of Black Hole}
Black hole could be formed by photons or relativistic particles. \cite{rf:HAW2,rf:CAR,rf:LEE}
 Here we try to
 derive the entropy of the black hole simply by adopting a few assumptions.
Denoting the mass and radius of black Hole by $M$ and $ r_g = 2GM/c^2 $, the energy of the 
particle $\Delta E$ which is constrained within this scale is estimated as
\begin{equation}
\Delta E \simeq \Delta pc \simeq \frac{ \hbar c }{ 2 \Delta x } \simeq k_B T , \hspace{2cm}
\nonumber
\end{equation}
 where $ k_B$ is the Boltzmann constant and we use the relation
$ \Delta E \simeq \Delta pc$ ,
\hspace{2pt} $\Delta x \Delta p \simeq \hbar / 2 , \hspace{2pt} \Delta E \simeq k_B T $. 
Now putting $ \Delta x \simeq c_1 r_g $, the temperature becomes
\begin{equation}
 T = \frac{ \hbar c }{ 2 c_1 k_B ( \frac{ 2GM }{ c^2 } ) } \simeq \frac{ \hbar c^3 }{ 4 c_1 GM k_B },
\nonumber
\end{equation}
which shows that temperature is proportional to the surface gravity $a=GM/r_g^2=c^4/GM$ as $T=\hbar a/c_1 k_B$.
If we take as $ c_1 = 2 \pi $, it becomes 
\begin{equation}
T =\frac{ \hbar a c }{ 2 \pi k_B } =\frac{ \hbar c^3 }{ 8 \pi GM k_B } \left ( = T_{BH} \right ), 
\nonumber
\end{equation}
which is the one that Hawking has derived. \cite{rf:HAW1}

If the black hole of mass $M$ is formed by the black body radiation of temperature $T$, the volume $V$, 
total photon number $N$ and total entropy $S$ are given by
\begin{equation}
V = \frac{ Mc^2 }{ \epsilon } =\frac{ 15 }{ \pi ^2 } \frac{ (8 \pi G)^4 M^5 }{ \hbar c^7 }, 
\nonumber
\end{equation}

\begin{equation}
N = nV = \frac{ 30 \zeta (3) }{ \pi ^4 } 
\frac{ 8 \pi G }{ \hbar c } M^2 \simeq \frac{ 240 }{ \pi ^3 } \frac{ G }{ \hbar c } M^2 ,
\nonumber
\end{equation}
\begin{equation}
\frac{ S }{ k_B } = \frac{ sV }{ k_B } = 
\frac{ 32 \pi }{3} \left ( \frac{ G M^2 }{ \hbar c } \right ) ,
\nonumber
\end{equation}
where the following  radiation density $ \epsilon $, 
number density $n$, and entropy density $s$ of the radiation temperature $T$  are used
\[ \epsilon = \tilde{a} T^4 \]
\begin{equation}
 n = \frac{ 2 \zeta (3) }{ \pi ^2 } \left ( \frac{ k_B T }{ \hbar c } \right )^3 = 0.244 \left ( \frac{ k_B T }{ \hbar c }
 \right )^3 ,\hspace{1cm}  
\nonumber
\end{equation}
\[ s = \frac{4}{3} \tilde{a} T^3 ,\]
\begin{equation}
\frac{n}{ \frac{s}{ k_B } } = \frac{ \frac{2}{ \pi ^2 } \zeta (3) }{ \frac{ \pi ^2 }{ 15 } \times \frac{4}{3} } = 
\frac{ 45 \zeta (3) }{ 2 \pi ^4 } \simeq 0.2776 ,
\end{equation}
being $ \tilde {a} = \pi ^2 k_B ^4 / ( 15 \hbar ^3 c^3 ) $ and $ \zeta (3)= 1.202 
 $ the 
radiation constant and zeta function. \cite{rf:LAN}

Using the surface of the black hole $A
= 4 \pi r_g ^2 =16 \pi   G^2 M^2 /c^4  $,
and considering the case not satisfying $ c_1 = 2 \pi $, the following relation is derived
\begin{equation}
\frac{S}{k_B} = \frac{2}{3} \frac{ c^3 A }{ G \hbar } = \frac{2}{3} \frac{A}{ \ell _p ^2 } =
 \frac{2}{3} \left( \frac{c_1}{2 \pi } \right ) \frac{ A }{ \ell _p ^2 } = \frac{1}{4} \left( \frac{c_1}{3 \pi /4} \right ) \frac{ A }{ \ell_p ^2 }= 
\frac{1}{4} \frac{ A }{ \ell_p ^2 }\hspace{2pt} ,\hspace{5pt}
\nonumber
\end{equation}
where $ \ell _p = \sqrt{ G \hbar / c^3 } $ is the Planck length.  The factor $c_1$ should be taken
 as $c_1 = \frac{3}{4} \pi $ when we consider entropy/

  In the above derivation, it is assumed implicitly that gravitational mass $M$ is equal to 
the proper mass $( M_p = \epsilon V / c^2 )$.  If we take the 
gravitational mass as $M= c_2 M_p $, the relation of the entropy to the surface 
becomes 

\[ \frac{S}{k_B} = \frac{ c_1 }{ 3 \pi c_2 } \frac{ A }{ \ell _p ^2 } 
= \frac{1}{4} \frac{ A }{ \ell _p ^2 } \hspace{2pt} ,\]
then we take 
\[\frac{ c_1 }{c_2} = \frac{ 3 \pi }{4} \hspace{2pt} , \]
so we get  $c_2 = \frac{8}{3}$, if we take $c_1 = 2 \pi$.
For the calculation of the neutron star, the gravitational mass is smaller than the proper mass $ c_2 < 1 $, 
and it is opposite for this case.  Even though the factor could not be well interpreted, it is derived the relation that entropy 
$S$ is proportional to the surface $A$.

  The problem is the volume $V$, which is much greater than the expected black hole volume
 $ V_{BH} \sim \frac{4}{3} \pi r_g ^3 $.  The ratio is
\begin{equation}
\frac{V}{V_{BH}} = \frac{V}{ \frac{4}{3} \pi r_g ^3 } = \frac{ \frac{15}{ \pi ^2 } 
\frac{ (8 \pi G)^4 M^5 }{ \hbar c^7 } }{ \frac{4}{3} \pi \left ( \frac{2GM}{c^2} \right )^3 } 
\nonumber
\end{equation}
\begin{equation}
 =\frac{ 45 }{ 4 \pi ^3 } \frac{ G }{ \hbar c } M^2 \frac{ ( 8 \pi )^4 }{ 2^3 } = 
\frac{45}{4} 8^3 \pi \left ( \frac{ M }{ m_p } \right )^2 = 90 \times 8^2 \pi  \left ( \frac{M}{m_p} \right )^2 ,
\nonumber
\end{equation}
then $ V \gg V_{BH} $ for  $M \gg m_p \hspace{4pt}$, where $m_p = \sqrt{ \hbar c / G } $ is the Planck mass.

  In the above derivation, we estimate the temperature from the uncertainty principle and assume that 
the black hole is formed from the gravitational force of the radiation of temperature $ T$. Then it is derived the 
relation that the entropy is proportional to the area of the black hole.
  In the normal star, the entropy is the order of $ S / k_B \sim N \sim 10^{57} \left ( M / M_{\odot} \right ) $, being 
$N$ the total number of particles, whereas the entropy of the black hole is order of 
$ S / k_B \sim A / ( 4 \ell _p ^2 ) \sim 10^{77} \left ( M / M_{\odot} \right )^2 $. This huge difference is the annoying 
problem of the entropy problem of the black hole. However it is interesting to note the following point.

 It is discussed the formation of black holes in the early time of the universe due to 
the large amplitude of fluctuations. \cite{rf:HAW2,rf:CAR,rf:LEE} From this point of view, the radiation energy should be 
enough to form the black hole within the volume  $V_{BH}$. Then 
the temperature $T_{ \gamma }$ must satisfy 
 \[ Mc^2 = \tilde {a} T_{ \gamma }^4 V_{BH} .\]

As  $V_{BH}\propto M^3$, it is derived $T_{ \gamma } \propto M^{- \frac{1}{2} }$.
The total entropy within the volume $V_{BH}$ for the temperature $T_{ \gamma }$ is given by
$S_{ \gamma } = \frac{4}{3} \tilde {a} T_{ \gamma }^3 V_{BH} $, so  it is derived 
$S_{ \gamma } \propto M^{ \frac{3}{2} } $. 
 If we put  $ S_{BH} / k_B = A / ( 4 \ell _p ^2 ) $, the ratio is 

\[ \frac{ S_{BH} }{ S_{ \gamma } } =
 \frac{3}{2} \left ( \frac{45 \pi }{2} \right )^{ \frac{1}{4} } \left ( \frac{M}{m_{pl}} \right )^{ \frac{1}{2} } = 
\frac{3}{8} \left ( \frac{ T_{ \gamma }}{ T_{BH} } \right ) .\]
Then the relation $S_{BH}T_{BH} \sim S_{ \gamma } T_{ \gamma } \sim U ( \sim Mc^2 ) $ is deduced.
The entropy has increased by $ T_{ \gamma } / T_{BH} ( \propto M^{ \frac{1}{2} } )$.
However it is difficult to understand in the microscopic process why the entropy has increased by such amount.
\cite{rf:HAR} 

\section{ Universe with $ \Lambda $ term}

The SNe Ia and WMAP observations confirm that our universe is now accelerating. For simplicity, we consider the universe with
 $\Lambda $ term as de Sitter space with metric \cite{rf:GIB}

\[ ds^2 = - \left( 1 - \frac{ \Lambda }{ 3 } r^2 \right ) c^2 dt^2 +
 \left ( 1 - \frac{ \Lambda }{ 3 } r^2 \right )^{-1} dr^2 + r^2 \left ( d \theta ^2 + \sin ^2 \theta d \phi ^2 \right ) .\]
 The characteristic point of this space is that there is the horizon with the radius of $ \ell _{ \Lambda } =
 \sqrt{ 3/ \Lambda }$ (de Sitter horizon).
Applying the uncertainty principle for this length, the energy $ \Delta E$ 
for the photon or particle is estimated as

\[ \Delta E \simeq \Delta pc \simeq \frac{ \hbar c }{ 2 \Delta x } \simeq k_B T . \]
Taking $ \Delta x = c_3 \ell _{ \Lambda } $, the temperature $T$ is given by
\[ k_B T = \frac{ \hbar c }{ 2 c_3 \ell _{ \Lambda } } \simeq 
\frac{ \hbar c }{ 2 c_3 \sqrt{3} } \sqrt{ \Lambda }. \]
When we put $ c_3 = \pi $ and the acceleration of the space as $a=c^2 \sqrt{\Lambda /3}$,
 it becomes the one $ k_B T = \hbar c \sqrt{\Lambda } / ( 2\pi \sqrt{3}) = a \hbar /(2\pi c) $ 
what Gibbons and Hawking have derived. \cite{rf:GIB}

  Because the cosmological constant is related to the vacuum energy density $\rho _{ \Lambda } $ as
\[ \rho _{ \Lambda } = \frac{ \Lambda c^2 }{ 8 \pi G } \]
the number density of the particle is given by
\[ n_{ \Lambda } \frac{ \Delta E }{ c^2 } = \rho _{ \Lambda } .\]
Assuming that particle energy is given by  $\Delta E = \hbar c / 2 c_3 \ell _{ \Lambda } $,
the number density $n_{\Lambda }$  becomes as

\[ n_{ \Lambda } = \frac{ \rho _{ \Lambda } c^2 }{ \Delta E } =
 \frac{ \Lambda c^4 }{ 8 \pi G } \frac{ 2 c_3 \ell _{ \Lambda } }{ \hbar c } = 
\frac{ 3 c_{3} }{ 4 \pi G } \frac{ c^3 }{ \hbar \ell _{ \Lambda } } . \]
Taking the volume of the universe $V$ as

\[ V = \frac{4}{3} \pi \ell _{ \Lambda } ^3 ,\]
the total number is given by
\[ N_{ \Lambda } = n_{ \Lambda } V = 
\frac{ 3 c_3 }{ 4 \pi G } \frac{ c^3 }{ \hbar \ell _{ \Lambda } } \frac{4}{3} \pi \ell _{ \Lambda } ^3 = 
c_3 \left ( \frac{ c^3 }{ G \hbar } \right ) \ell _{ \Lambda } ^2 = c_3 \frac{ \ell _{ \Lambda } ^2 }{ \ell _p ^2 } .\]

If we assume the particle as Bose particle such as photon, the total entropy is proportional to the total number as
(see eq.(2.1) ; $  N /( S / k_B ) = 0.2776 $  )

\[ \frac{S}{k_B} \sim \frac{ N }{ 0.2776 } \sim 4 c_3 \frac{ \ell _{ \Lambda } ^2 }{ \ell _p ^2 }. \]
Using the area of the de Sitter horizon $A= 4 \pi \ell _{ \Lambda } ^2 $, it is expressed as
\[ \frac{S}{k_B} \sim \frac{c_3}{\pi } \frac{A}{ \ell _p ^2 } ,\]
where the entropy is proportional to the horizon area $A$.
If we take $c_3= \pi / 4 $, it becomes  $ S / k_B = A / ( 4 \ell _p ^2 )$ which is derived by Gibbons and Hawking. \cite{rf:GIB}

   It must be noted that the equation of state for the matter corresponding to cosmological constant is 
$p= - \rho c^2 $, which is different from the photons. It might be some scalar fields.
The problem of the above derivation is that the energy density due to the cosmological constant is very 
different from the inferred radiation energy density $ \epsilon _{ \gamma } = \tilde {a} T^4 $ of temperature $T$. 
The ratio of $\rho _{ \Lambda }$  to $ \rho _{\gamma } $ is

\begin{eqnarray}
\frac{ \rho _{ \Lambda } }{ \rho _{ \gamma } } 
= \frac{ \rho _{ \Lambda } }{ \frac{ \epsilon _{ \gamma } }{ c^2 } } =
\frac{ \Lambda c^2 }{ 8 \pi G } \frac{ c^2 }{ \tilde {a} T^4 }
\simeq \frac{ c^3 }{ 8 \pi G \hbar } \frac{15}{ \pi ^2 \Lambda } \left ( 2 c_3 \sqrt{3} \right )^4 
\simeq \frac{90}{ \pi ^3 } c_3 ^4 \left ( \frac{ \ell _{ \Lambda } }{ \ell _p } \right )^2 .
\end{eqnarray}
As $ \ell _{ \Lambda } \gg \ell _p $, the relation $ \rho _{ \Lambda } \gg \rho _{ \gamma } $ is derived.

   If we take that the energy density $ \rho _{pl} $ of the vacuum due to the zero-point energy is  
\cite{rf:WEI,rf:PAD2}

\[ \rho_{pl} \simeq \frac{ m_{pl} }{ \ell _p ^3 } \simeq \frac{ \hbar }{ \ell _p ^4 c } \hspace{15pt} ,\]
and the ratio with $ \rho _{ \Lambda }$ is

\[ \frac{ \rho _{ \Lambda } }{ \rho _{pl} } 
= \frac{ \frac{ \Lambda c^2 }{ 8 \pi G } }{ \frac{ \hbar }{ \ell _p ^4 c } } =
 \frac{ 3 }{ 8 \pi } \left ( \frac{ \ell _p }{ \ell _{ \Lambda } } \right )^2 . \]
If the above relations are considered, the following relation is derived 
\[ \frac{ \rho _{ \Lambda }}{ \rho _{ \gamma } } \times \frac{ \rho _{ \Lambda } }{ \rho _{pl} } 
\simeq \frac{90}{ \pi ^2 } c_3 ^4 \left ( \frac{ \ell _{ \Lambda } }{ \ell _p } \right )^2 \times 
\frac{3}{ 8 \pi } \left ( \frac{ \ell _p }{ \ell _{ \Lambda } } \right )^2 \simeq {\it O} (1), \]
and it is induced

\[ \rho _{ \Lambda } \simeq \sqrt{ \rho _{ \gamma } \rho _{pl} } .\]
It means that the energy density   $ \rho _{ \Lambda } $ of the cosmological constant is the geometrical mean 
of the radiation energy density 
$ \rho _{ \gamma }$ of the temperature $T$ originated from the uncertainty principle and the vacuum 
energy density $ \rho _{pl} $ from the zero-point energy.

\section{Uniformly accelerating coordinate}

It has been said as Unruh Effect that to the observer in the uniformly accelerating coordinate
it seems that he or she is in a bath of blackbody radiation at the temperature $T$  which is related 
to the acceleration $ \kappa (=a)$ as
 $k_B T = \hbar \kappa  / ( 2 \pi c ) $. \cite{rf:MAT}

There is a characteristic length $\ell _{ \kappa } =  c^2 / \kappa $ due to the acceleration $ \kappa $.
Applying the uncertainty principle to this length, the energy $\Delta E$ of the particle is given by 

\[ \Delta E \simeq \Delta pc \simeq \frac{ \hbar c }{ 2 \Delta x } \simeq \frac{ \hbar c }{ 2 c_4 \ell _{ \kappa } }
 \simeq \frac{ \hbar \kappa }{ 2 c_4 c } \simeq k_B T ,\]
where we put $ \Delta x = c_4 \ell _{ \kappa } $. The relation between the temperature and the acceleration 
$ \kappa $ is  given by

\[ k_B T \simeq \frac{ \hbar \kappa }{ 2 c_4 c } .\]
If we take $c_4 = \pi $, the result is the same derived by Unruh. \cite{rf:UNR}

In the following we consider the relation of this temperature to the entropy as
\[ k_B S = \frac{ A }{ 4 \ell _p ^2 }. \]
One way of the derivation is to assume that the acceleration $ \kappa $ is the gravitational acceleration
 by the mass $M$,
which is composed of the blackbody radiation of temperature $T$. If we take the volume of the considering region 
as $V$, the mass is given as
\[ M = \frac{ \tilde {a} T^4 }{ c^2 } V .\]
The acceleration $ \kappa $ due to this mass is given by

\[ \kappa = \frac{ GM }{ V^{ \frac{2}{3} } } = \frac{ G \tilde {a} T^4 }{c^2} \frac{V}{ V^{ \frac{2}{3} } } = 
\frac{ G \pi ^2 }{15} \frac{ k_{B}^4 T^4 }{ \hbar ^3 c^3 } \frac{ V^{ \frac{1}{3} } }{c^2} .\]
Here we use the relations $ k_{B} T = \hbar c / ( 2 c_{4} \ell _{ \kappa } ) $, the above equation becomes 

\[ \kappa =\frac{ c^2 }{ \ell _{ \kappa } } = \frac{ \pi ^2 }{15} \frac{1}{ 16 c_{4} ^4 } \frac{ G \hbar c }{ c^2 } 
\frac{ V^{ \frac{1}{3} } }{ \ell _{ \kappa }^4 }.\] 
Then  the region size is given by

\[ V^{ \frac{1}{3} } = \frac{ 15 \times 16 c_{4} ^4 }{ \pi ^2 } \frac{ \ell _{ \kappa } ^3 }{ \ell _p ^2 } .\]

As the total entropy is  $ S = \frac{4}{3} \tilde {a} T^3 V $, the entropy per surface $ S / V^{\frac{2}{3}} $ is given by

\[ \frac{ \frac{S}{k_B} }{ V^{ \frac{2}{3} } } = \frac{4}{3} \frac{ \pi ^2 }{15} 
\frac{ k_B ^3 T^3 }{ \hbar ^3 c^3 } V^{ \frac{1}{3} } = \frac{4}{3} \frac{ \pi ^2 }{15} \frac{ \frac{15}{ \pi ^2 } \times 
16 c_4 ^4 \frac{ \ell _{ \kappa } ^3 }{ \ell _p ^2 } }{ 8 c_4 ^3 \ell _{ \kappa } ^3 } =
 \frac{8}{3} c_4 \frac{1}{ \ell _p ^2 } .\]
If we take  $ c_4 = \frac{3}{32} $ and  $ A = V^{ \frac{2}{3} } $, the following relation is derived

\[ \frac{ \frac{S}{ k_B }}{A} = \frac{1}{ 4 \ell _p ^2 } ,\]
which means that the entropy per surface is $ 1 / ( 4 \ell _p ^2 ) .$

Even though we consider by the order estimation, it is derived the relation $ S / ( k_B A ) = 1 / ( 4 \ell _p ^2 ) $, 
which
is the common relation of the entropy to the surface of the event horizon of the black hole and de Sitter universe.

\section{Consideration on higher dimensions}

In this section, we consider the above 3 cases in the 4+n dimensions $( n \geq 0)$.
 \subsection{Spherical symmetric case}

The Schwarzschild metric for 4+n dimension is given by

\[ ds^2 = - h(r) c^2 dt^2 + h(r) ^{-1} dr^2 + r^2 d \Omega_{2+n} ^2 ,\]
where  $ d \Omega _{2+n} ^2 $ is the surface metric of unit sphere in 3+n dimensions, so  $h(r)$ is written as

\[ h(r) = 1 - \left( \frac{ r_H }{ r } \right)^{n+1} =
 1 - \frac{16 \pi }{ n+2 } \frac{ G(n) }{ A_{n+2} } \frac{ M_{BH} }{ c^2 } \frac{1}{r^{n+1}}. \]
$G(n)$ is the gravitational constant in 4+n dimension and $ A_{n+2} $ is the surface area of unit sphere as 

\[ A_{n+2} = \frac{ 2 \pi ^{\frac{n+3}{2}} }{ \Gamma ( \frac{n+3}{2} ) }, \]
where it becomes $A_2 = 4 \pi $ for $n=0$, because $\Gamma (3/2) = \sqrt{ \pi }/{2} $.

The Planck mass, length, and time are given by

\[ m_{pl} (n) = \left ( \frac{ \hbar ^{n+1} c^{1-n} }{ G(n) } \right )^{ \frac{1}{2+n} } , \hspace{0.3cm}
\ell _{pl} (n) = \left ( \frac{ G(n) \hbar }{ c^3 } \right )^{ \frac{1}{2+n} } ,  \hspace{0.3cm}
t_{pl} (n) = \left ( \frac{ G(n) \hbar }{ c^{ 5+n } } \right )^{ \frac{1}{2+n} } ,\]
respectively.  So $ r_H$ is expressed as 

\[ r_H = \frac{1}{ \sqrt{ \pi } } \frac{1}{ m_{pl} (n) } \frac{ \hbar }{c}  
\left( \frac{ M_{BH} }{ m_{pl} (n) } \right) ^{ \frac{1}{n+1}} 
\left( \frac{ 8 \Gamma (\frac{n+3}{2}) }{ n+2 } \right)^{ \frac{1}{n+1} } .\]

Due to the uncertainty principle, the energy of the wave which is confined to the black hole 
is given as
\[ \Delta E \sim \Delta pc \sim \frac{ \hbar c }{ \Delta r } 
\sim kT \Longrightarrow kT \sim \frac{ \hbar c }{ r_H } .\]
The radiation energy density $ \epsilon (T) $ and the entropy density  $s(T)$  are given by

\[ \epsilon (T) = \frac{ A_{n+2} }{ (2 \pi )^{ n+3 } } k_B T(n+2) 
\left ( \frac{ k_B T }{ \hbar c } \right )^{ n+3 } \Gamma (n+4) \zeta (n+4) \sim (k_B T)^{n+4} \]
and 

\[ s(T) = \frac{(n+4)(n+2) }{n+3} \frac{ A_{n+2} }{ (2 \pi )^{3+n} } k_B 
\left ( \frac{k_B T}{\hbar c} \right)^{n+3} \Gamma (n+4) \zeta (n+4) \sim (k_B T)^{n+3}, \]
respectively.  Then  the volume could be estimated as 

\[ V = \frac{ Mc^2 }{ \epsilon } \propto \frac{ Mc^2 }{ k_B T \left ( \frac{ k_B T }{ \hbar c } \right )^{n+3} } \]
and the total entropy becomes

\[ S = sV = k_B \left ( \frac{ k_B T }{ \hbar c } \right )^{n+3} \frac{ Mc^2 }{ k_B T 
\left ( \frac{ kT }{ \hbar c } \right )^{n+3} } \sim \frac{ Mc^2 }{T} \sim
 \frac{ k_B r_H Mc^2 }{ \hbar c } \]
\[\sim \frac{ k r_H }{ \hbar c }
 \left ( \frac{ r_H }{ \ell _{pl} (n) } \right )^{n+1} m_{pl} c \sim k \left ( \frac{ r_H }{ \ell _{pl} } \right )^{n+2} ,\]
which means that the entropy in the 4+n dimension is proportional to the 2+n dimensional  hyper surface  in n+3 space dimension.
Therefore the entropy of black hole is proportional to the surface.

\subsection{ de Sitter space}

The metric is given by

\[ ds^2 = - \left ( 1 - \frac{ \Lambda }{ 3 } r^2 \right ) c^2 dt^2 + 
\left ( 1 - \frac{ \Lambda }{ 3 } r^2 \right )^{-1} dr^2 + r^2 d \Omega _{n+2} ^{2} .  \]
Taking $\ell _{ \Lambda } = ( 3/ \Lambda )^{ \frac{1}{2} } $, the temperature of the universe $T$ is 

\[ \Delta E \sim \Delta pc \sim \frac{ \hbar c }{ \ell _{ \Lambda } } \sim k_B T .\]
The cosmological constant $\Lambda $ is related to the density as

\[ \rho (n) = \frac{ \Lambda c^2 }{ 8 \pi G(n) }, \]
then the particle density $n_{ \Lambda }$ is given by

\[ n_{ \Lambda } \frac{ \Delta E }{ c^2 } \sim \rho _{ \Lambda } \sim \frac{ \Lambda c^2 }{ 8 \pi G(n) } .\]
Taking $ V \sim \ell _{ \Lambda }^{3+n} $, the total particle number becomes as 

\[ N_{\Lambda } =
 n_{\Lambda } V \sim \frac{ \Lambda c^4 }{ 8 \pi G(n) } \frac{ \ell _{ \Lambda } }{ \hbar c } \ell _{ \Lambda } ^{ 3+n } 
\sim \frac{ \ell _{ \Lambda } ^{2+n} }{ \ell _{pl} ^{2+n} } = \left ( \frac{ \ell _{ \Lambda }}{ \ell _{pl} } \right )^{2+n}. \]
As the total enttropy $S$ is proportional to $N_{ \Lambda }$,  the following relation  is derived,

\[ S \propto \left ( \frac{ \ell _{ \Lambda }}{ \ell _{pl} } \right )^{2+n} ,\]
which means that the total entropy is proportional to the horizon surface.

\subsection{Rindler space}

Denoting the constant acceleration as $\kappa $,  the characteristic length becomes $ \ell _{ \kappa } = c^2 / \kappa $
 and the temperature $T$ is estimated, 

\[ \Delta E \sim \Delta pc \sim \frac{ \hbar c }{ 2 \ell _{ \kappa } } \sim k_B T .\]
We tentatively assume that the acceleration is due to the gravity of mass $ M$.  Taking the volume $V=L^{3+n}$ and 
the mass $M$ as  

\[ M = \frac{ \epsilon }{ c^2 } V \sim \frac{ k_B T }{ c^2 } \left ( \frac{ k_B T }{ \hbar c } \right )^{3+n} L^{3+n} ,\]
 the acceleration $ \kappa $ is given by

\[ \frac{c^2}{ \ell _{ \kappa } } \sim \kappa = \frac{ G(n) M }{ L^{2+n} } \sim \frac{ G(n) }{ L^{2+n} }
 \frac{ k_B T }{ c^2 } \left ( \frac{ k_B T }{ \hbar c } \right )^{3+n} L^{3+n} \sim G(n) 
\left ( \frac{ 1 }{ \ell _{ \kappa }} \right )^{4+n} \frac{ \hbar }{ cL } .\]
Then the length $L$ becomes

\[ L = \frac{ c^3 }{ G(n) \hbar } \ell _{ \kappa } ^{3+n} = 
\left ( \frac{ \ell _{ \kappa } }{ \ell _{pl} } \right )^{2+n} \ell _{ \kappa }.\]
As the total entropy is $S=sV=sL^{3+n}$, the entropy per unit surface is 

\[ \frac{ S }{ L^{2+n} } \sim sL \sim k_B 
\left ( \frac{1}{ \ell _{ \kappa }} \right )^{2+n} \ell _{ \kappa } 
\left ( \frac{ \ell _{ \kappa }}{ \ell _{pl}} \right )^{2+n} \sim k_B \left ( \frac{1}{ \ell _{pl}} \right )^{2+n} ,\]
or it is written as
\[ \frac{S}{ \left ( \frac{L}{ \ell _{pl} } \right )^{2+n} } \sim k_B . \]
It is also the natural extension of the constant of entropy per surface to 4+n dimension case.

\section{Conclusions and discussion}

By using the uncertainty principle, it is estimated the temperature of the black hole, de Sitter space, and Rindler 
space are commonly expressed by the acceleration $a$ as $T=a/2\pi$. Considering 
the black body radiation of this temperature, it is derived the common relation 
that the entropy is proportional to the surface of the event horizon for each case as $S=A/4$.  
It is also derived the curious 
relation that the energy density related to cosmological constant $\rho _{ \Lambda } = 3 \Lambda / ( 8 \pi G ) $ is 
the geometrical mean of 
the radiation density $\rho _{ \gamma }$ and the vacuum energy density $ \rho _{pl}$ as 
$\rho _{ \Lambda } = \sqrt { \rho _{ \gamma }\rho _{pl} }$.

  There are many points to be investigated further in the above derivation.  It is not treated here that the problems about the general second law of thermodynamics, 
black hole evaporation and the surface increase of the cosmological event horizon with $ \Lambda $ term. 
It is interesting to investigate these problems under these simplified assumptions.

$ \vspace{5pt} $


\begin{thebibliography}{99}
\bibitem{rf:BEK}
 J. D. Bekenstein, Phys. Rev. {\bf D7}(1973), 2333.
\bibitem{rf:JAC} 
T. Jacobson and R. Parentani, Found. Phys. {\bf 33}(2003), 323-348.
\bibitem{rf:HAW1} 
S. W. Hawking, Commun. Math. Phys.{\bf 43}(1975), 199.
\bibitem{rf:BEK2} 
J. D. Bekenstein, Phys. Rev. {\bf D7}(1979), 2333-2346.
\bibitem{rf:BIR} 
N. D. Birrell and P. C. W. Davies, {\it Quantum Fields in Curred Space}, (Cambridge University Press, 1982).
\bibitem{rf:CARL} 
 S. Carlip, arXiv:gr-gc/0601041.
\bibitem{rf:HOO} 
't Hooft, G., arXiv : gr-qc/9310026.
\bibitem{rf:MAR} 
J. M. Maldacena, Adv. Theor. Math. Phys. 2(1998), 231.
\bibitem{rf:POL}
J. Polchinski, \textit{String Theory} (Cambridge University Press, 1998)
\bibitem{rf:SCA} 
 F. Scardigli, Nuovo Cimento, {\bf 110B}(1995), 1029.
\bibitem{rf:PER} 
 S. Perlmutter et al., Ap. J. {\bf 517}(1999), 565. 
\bibitem{rf:RIE} 
 A. G. Riess et al., Ap. J.  {\bf 607}(2004), 665.
\bibitem{rf:SPE} 
 D. N. Spergel et al., Ap. J. S. {\bf 148}(2003), 175.
\bibitem{rf:PAD1} 
T. Padmanabhan, Mod. Phys. Lett.  {\bf A17}(2002) 923-942.
\bibitem{rf:HAW2} 
S. W. Hawking, Mon. Not. R. astr. Soc. {\bf 152}(1971), 75.
\bibitem{rf:CAR}
B. J. Carr and S. W. Hawking, Mon. Not. R. astr. Soc. {\bf 168}(1974), 399.
\bibitem{rf:LEE}
H. K. Lee, Phys. Rev.  {\bf D66} (2002), 063001.
\bibitem{rf:LAN} 
L. D. Landau and E. M. Lifshitz, {\it Statistical Physics}, Reed Edu. and Pro. Pub.Ltd. Oxford(1980).
\bibitem{rf:HAR}
T. Hara et al.,  arXiv: gr-qc/0503105.
\bibitem{rf:GIB} 
G. W. Gibbons and S. W. Hawking, 
Phys. Rev. {\bf D15}(1977) 2738-2751.
\bibitem{rf:WEI} 
 S. Weinberg, Rev. Mod. Phys. {\bf 61}(1989), 1.
\bibitem{rf:PAD2}
 T. Padmanabhan, 
Class. Quant. Grav. {\bf 19}(2002), 5387-5408.
\bibitem{rf:MAT}
 G. E. A. Matsas and D. A. T. Vanzella,
Int. J. Mod. Phys. {\bf D11}(2002), 1573-1578.
\bibitem{rf:UNR} 
 W. G. Unruh, Phys. Rev. {\bf D14}(1976), 870.
\end{thebibliography}
\end{document}